\documentstyle[11pt]{article}
\input epsf
\def\thefootnote{\fnsymbol{footnote}}

\newcommand{\nn}{\nonumber}

\newcommand{\ri}{\right}
\newcommand{\lf}{\left}

\newcommand{\ep}{\varepsilon}
\newcommand{\eq}{\begin{equation}}
\newcommand{\en}{\end{equation}}
\newcommand{\bea}{\begin{eqnarray}}
\newcommand{\eea}{\end{eqnarray}}

\newcommand{\ba}{\begin{array}}
\newcommand{\ea}{\end{array}}

\newcommand{\CC}{{\hbox{\rm C\kern-0.5em{$\sf I$}}}}
\newcommand{\II}{\hbox{{\rm l{\hbox to 1.5pt{\hss\rm l}}}}}
\newcommand{\RR}{{\hbox{$\rm\textstyle I\kern-0.2em R$}}}
\newcommand{\ZZ}{{\hbox{$\sf\textstyle Z\kern-0.4em Z$}}}

\newcommand{\resection}[1]{\setcounter{equation}{0}\section{#1}}

\newcommand{\NP}[1]{Nucl.\ Phys.\ {\bf #1}}
\newcommand{\PL}[1]{Phys.\ Lett.\ {\bf #1}}

\newcommand{\CMP}[1]{Comm.\ Math.\ Phys.\ {\bf #1}}
\newcommand{\CPC}[1]{Computer Phys.\ Comm.\ {\bf #1}}
\newcommand{\PR}[1]{Phys.\ Rev.\ {\bf #1}}
\newcommand{\PRL}[1]{Phys.\ Rev.\ Lett.\ {\bf #1}}

\newcommand{\PTPS}[1]{Prog.\ Theor.\ Phys.\ Supp.\ {\bf #1}}

\newcommand{\IJMP}[1]{Int.\ J.\ Mod.\ Phys.\ {\bf #1}}

\newcommand{\AlBZ}{Al.B.Zamolodchikov}

\newcommand{\iintd}{\int^{\infty}_{-\infty}\! d\theta}

\newcommand{\MR}{M\! R}

\newcommand{\fract}[2]{{\textstyle\frac{#1}{#2}}}

\newcommand{\halft}{\fract{1}{2}}
\newcommand{\prtial}{\frac{\partial}{\partial\theta}}

\newcommand{\opnup}[1]{\renewcommand{\\}{\\[50 pt]}}
\renewcommand{\bar}{\overline}
\newcommand\IM{^{\rm IM}}
\newcommand\re{{\rm Re}}
\newcommand\im{{\rm Im}}
\newcommand\Arg{{\rm Arg}}
\newcommand\LYM{{\cal M}(2/5)}
\begin{document}
\begin{titlepage}
\vskip 0.5cm
\begin{flushright}
DTP-96/29 \\
July 1996 \\
hep-th/9607167
\end{flushright}
\vskip 1.5cm
\begin{center}
{\Large {\bf Excited states by analytic continuation}} \\[5pt]
{\Large {\bf of TBA equations } }
\end{center}
\vskip 0.9cm
\centerline{Patrick Dorey and Roberto Tateo}
\vskip 0.6cm
\centerline{\sl Department of Mathematical Sciences,  
}
\centerline{\sl  University of Durham, Durham DH1 3LE, 
England\,\footnote{e-mail: {\tt P.E.Dorey@durham.ac.uk,
Roberto.Tateo@durham.ac.uk}} }
\vskip 1cm
\begin{abstract}
\vskip0.2cm
\noindent
We suggest an approach to the problem of finding integral equations 
for the excited states of an integrable model, starting from the
Thermodynamic Bethe Ansatz equations for its ground state. The idea
relies on analytic continuation through complex values of the coupling
constant, and an analysis of the monodromies that the equations
and their solutions undergo. For the scaling Lee-Yang model, we find
equations in this way for the one- and two- particle states in the
spin-zero sector, and suggest various generalisations. Numerical
results show excellent agreement with the truncated conformal space
approach, and we also treat some of the ultraviolet and infrared
asymptotics analytically.
\bigskip

\end{abstract}
\end{titlepage}

\setcounter{footnote}{0}
\def\thefootnote{\arabic{footnote}}

\resection{ Introduction }

The Thermodynamic Bethe Ansatz (TBA)~\cite{Zb} has 
proved to be a very useful tool
in the study of integrable two-dimensional field theories. The
finite-volume ground state energy can now be found for many models,
expressed in each case in terms of the solution of a set of nonlinear
integral equations. These turn out to be vulnerable to both numerical and
analytical attack, and yield a large amount of nontrivial information. 
Given these successes, it is natural to hope that similar equations might
describe the remaining energy levels. However, the derivation of the TBA
equations relies on arguments which perforce single
out the ground state, and, save for a few states which become
degenerate with the ground state in large volumes~\cite{KMc,Mc,Fc}, the
desired generalisation has proved elusive.

In this paper we propose to sidestep the problem by returning to the old
idea that one can move between energy levels by analytic continuation in a
suitable parameter. In 0+1 dimensions, the quantum-mechanical problem,
this was seen most spectacularly in the analysis of the quantum
anharmonic oscillator performed by Bender and Wu~\cite{BWa}. 
Somewhat simpler in analytic structure, but
more relevant to the following, is the finite volume spectrum of the Ising
field theory in 1+1 dimensions. If the `spatial' dimension is rolled up into 
a circle of circumference $R$, then the ground state energy (the lowest
eigenvalue of the infinitesimal transfer matrix in the `time'
direction) can be written as~\cite{KMb}
\eq
E\IM_0(M,R)=E\IM_{\rm bulk}(M,R)-{\pi\over 6R}c\IM_0(\MR)~,
\label{elgar}
\en
where $M$ the inverse of the bulk correlation length,
$E\IM_{\rm bulk}(M,R)$ contains the (as it happens logarithmically
divergent) bulk contribution, and
\bea
c\IM_0(r)&=&{1\over 2} -{3r^2\over 2\pi^2}\lf[\log{1\over r} + {1\over 2}
 + \ln\pi -\gamma_E\ri]\nn\\
&&~+{6\over\pi}\sum_{k=1}^{\infty}\lf(\sqrt{r^2+(2k{-}1)^2\pi^2}
 -(2k{-}1)\pi -{r^2\over 2(2k{-}1)\pi}\ri)
\label{bach}~~~\\ \nn
\eea
with $r=\MR$ and
$\gamma_E= 0.57721566..$. Apart from the logarithmic singularity at $R=0$,
this exhibits a series of square root branch cuts, evenly spaced along
the imaginary axis. Suppose that $R$ is continued into the complex
plane along a path enclosing the singularities at $k=k_1,k_2\dots k_n$. Then 
on its return to the real axis, $E\IM_0(M,R)$ has been replaced by
\eq
E\IM_{k_1,k_2\dots k_n}(M,R)=E\IM_0(M,R)
+{2\over R}\sum_{i=1}^n\sqrt{r^2+(2k{-}1)^2\pi^2}~,
\en
an excited state with ultraviolet scaling dimension 
$\sum_i(2k_i{-}1)$. This covers all of the symmetrical
descendants of the primary
fields $I$ and $\ep$ in the spin-zero sector. To find descendants
of the spin field $\sigma$, the same process can be repeated, but this
time
starting from the lowest excited state in the low-temperature regime, 
which degenerates with $E_0$
in infinite volume and is accessible as the ground state with twisted
boundary conditions (see for example~\cite{Fc}).

In more general situations explicit expressions such as 
(\ref{elgar}), (\ref{bach})
are not available. Even for integrable perturbations of
conformal field theories, the best one can do is to express the ground
state energy in terms of the solutions $\ep_a(\theta)$ to a set
of TBA 
equations. The simplest case is the scaling Lee-Yang model, or SLYM.
This field theory, a perturbation of the non-unitary minimal model 
$\LYM$ by its unique relevant operator $\varphi$, has the action
\eq
{\cal A}_{\rm SLYM}={\cal A}_{\LYM}+ \imath\lambda\int\varphi(x) d^2x~.
\label{arnold}
\en
With $M(\lambda)=(2.642944\dots)\lambda^{5/12}$~\cite{Fb} 
and $r=M(\lambda)R$, 
the single TBA equation reads~\cite{Zb}
\eq
\ep(\theta)=r\cosh\theta-\phi{*}L(\theta)~,
\label{holst}
\en
where
\eq
L(\theta)= 
\log\lf(1+e^{-\ep(\theta)}\ri)~,~~f{*}g(\theta)={1\over 2\pi}
\iintd' f(\theta-\theta')g(\theta)~,
\en
and
\eq
\phi(\theta)=-\imath\prtial\log S(\theta)~,~~S(\theta)=
{\sinh(\theta)+\imath\sin(\pi/3)\over\sinh(\theta)-\imath\sin(\pi/3)}~.
\en
The function $S(\theta)$ is the $S$-matrix of the single neutral
particle in the model~\cite{CMa}, 
and has the `$\phi^3$' bootstrap property, that
$S(\theta{-}\imath\pi/3)S(\theta{+}\imath\pi/3)=S(\theta)$.
In terms of these quantities, the ground state energy is
\eq
E_0(\lambda,R)=E_{\rm bulk}(\lambda,R) -{\pi\over 6R}c_0(r)
\en
with
\eq
E_{\rm bulk}(\lambda,R)={-M(\lambda)^2\,\over 4\sqrt{3}}R~,~~~~
c_0(r)= {3\over \pi^2}\iintd\,
 r\cosh\theta L(\theta)~.~~
\label{britten}
\en
At first sight (\ref{holst})--(\ref{britten}) are very different from
(\ref{elgar}), (\ref{bach}), and it is not clear that analytic continuation
will be practicable. To get a clue as to how to proceed, return to
(\ref{bach}) and consider its alternative
integral representation:
\eq
c\IM_0(r)= {3\over \pi^2}\iintd\,
 r\cosh\theta\log\lf(1{+}e^{-r\cosh\theta}\ri)~.
\en
As $r$ moves into the complex plane, a singularity in $c\IM$ might be
expected whenever $1{+}e^{-r\cosh\theta_0}=0$ for some real $\theta_0$.
However, deforming the contour of integration away from the real axis near
$\theta_0$ shows that such impressions are generally deceptive. This
man\oe uvre
would only fail if two singularities were to approach the real axis from
opposite sides, trapping the contour. This gives rise to a so-called `pinch
singularity', usually a branch point. If $r$ is
continued along some path encircling the critical value, the two
$\theta$-singularities (in this case, singularities in 
$\log(1{+}e^{-r\cosh\theta})$) execute a  
little dance in the complex plane (here, they just swap over), after which
the contour has become tangled up. Undoing the tangle gives rise to the
discontinuity across the cut. All of this is very clearly explained in
chapter 2 of the book by Eden {\it et al}~\cite{ELOP}, and
here we simply remark that
it is an instructive exercise to recover the results already quoted by the
use of such methods.

For more general TBA systems, life is complicated by the replacement of
$r\cosh\theta$ by $\ep(\theta)$, a function which may itself be subject to
nontrivial monodromies. Furthermore, $\ep(\theta)$ is not known explicitly
even for the ground state. Fortunately, only qualitative, `topological',
information about the movement of singularities is needed in order to deduce
the modified TBA equations, and for this numerical work will suffice.

\resection{ One-particle states}

\begin{figure}[htbp]
\vspace{-0.9cm}
\hspace{1.cm}
\epsfxsize=9.5cm
\epsfysize=9.5cm
\epsfbox{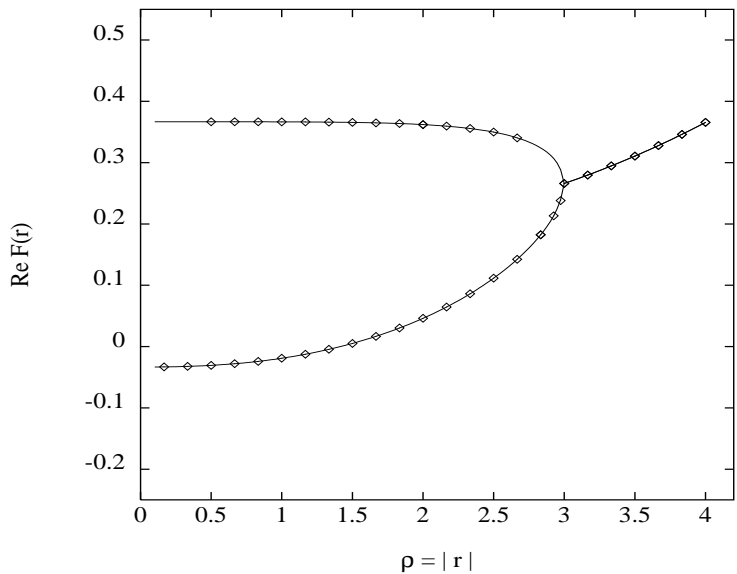}
\vspace{-0.3cm}
\caption{ \leftskip=1.3cm  Two solutions to the basic TBA equation on
the negative-$\lambda$ line (points) versus TCSA data (continuous 
lines) \rightskip=0.7cm }  
\label{neglambdafig}
\end{figure}

Both the truncated conformal space approach (TCSA)~\cite{YZa}, and the
numerical extrapolation of TBA results found for $\lambda\in\RR^+$,
have led to the conclusion that the ground state energy of the SLYM has a
square root singularity at $R(-\lambda)^{5/12}\approx 1.1325$. In fact, this
can also be seen directly from the initial TBA system 
(\ref{holst}). For this it is convenient to adsorb the bulk term into
$c_0(r)$, and work instead with the ground-state scaling 
function~$F_0(r)$:
\eq
E_0(\lambda,R)={2\pi\over R}F_0(r)~,~~F_0(r)=-{r^2\over
8\sqrt{3}\pi}-{1\over 12}c_0(r)~.
\en
The function $F_0$ thus defined is expected to be a regular function
of $r^{12/5}$~\cite{Zf}, well-suited to analytic continuation. Negative 
values of $\lambda$ put $r$ on the ray $r=\rho e^{5\pi\imath/12}$,
$\rho\in\RR^+$. A suitably-damped numerical iteration of the TBA 
equation (\ref{holst}) is convergent along this line, and the
resulting ground state scaling function turns out to be real out to
$\rho=\rho_0\approx 2.99315$, there being clear evidence for 
a square-root
singularity at this point. This is shown in the lower set of points
in
figure~\ref{neglambdafig}, and matches in 
all aspects the TCSA results found earlier by
Yurov and Zamolodchikov~\cite{YZa}. We conclude that the TBA is able to
provide reliable (and, indeed, highly accurate) information even away
from real values of $r$.

\setlength{\unitlength}{1.mm}

Next, the idea is to circle around the singularity, in the hope of
picking up the next branch of the function -- the first excited state.
Analytic continuation is straightforwardly implemented when solving
the TBA iteratively, by varying $r$ step by step and at each new value
of $r$ taking the initial iterate $\ep^{(0)}$ to be the
final iterate $\ep^{(n)}$ of the previous step. The steps in $r$ must
not be too large, or the solution being tracked may be lost. 
Also, one must guard against numerical instabilities which are purely
artifacts of the iteration scheme. We adopted the simplest possible
option, iterating as
\eq
\ep^{(n{+}1)}(\theta)
=a\lf[ r\cosh\theta-\phi{*}L^{(n)}(\theta)\ri]
+(1{-}a)\ep^{(n)}(\theta)~,
\label{tippett}
\en
and finding empirically that values of $a$ between $0.5$ and $0.05$
(depending on the value of $r$) gave optimal results. Stability rather
than efficiency turns out to be the key issue, and it is worth seeking
a more sophisticated approach. Nevertheless, (\ref{tippett}) was
adequate for the current work.

Of more profound import are the changes needed should a singularity of
$L(\theta)$ cross the real axis. In such a situation, further
analytic continuation results in a function $\ep$ which solves
a modified TBA equation, with the
contour of integration diverted away from from its original track
long enough to avoid
the singularity. To check for this, we must locate the complex
zeroes and poles
in $z(\theta)\equiv 1+e^{-\ep(\theta)}$, 
and be on our guard whenever any of them 
venture too close to the real axis. This is easily done, at least
numerically -- once $\ep(\theta)$ is known along the real axis (or
along some more general contour), equation~(\ref{holst}) provides an 
integral representation which can be used to reconstruct the function
everywhere. The only point to watch is that the singularities of
$\phi$ at $\pm\imath\pi/3$ and beyond
necessitate the introduction of extra terms
when $\ep(\theta)$ is continued beyond the strip
$-\imath\pi/3<\im\theta<\imath\pi/3$. 

Near the real axis at
large values of $|\re(\theta)|$, the $r\cosh\theta$ term 
comes to dominate (\ref{holst}), and so two 
series of zeroes in $z(\theta)$ are always seen, approaching
the two half lines
$\pm\{\theta\in\CC:\re\theta>0,\im\theta=[\pi/2-\Arg(r)]\}$. 
However if $r$ lies on
the initial segment of the negative-$\lambda$ line, 
$\Arg(r)=5\pi/12$ and $0<|r|<\rho_0$, we found evidence 
that, for the solution just discussed,
a stronger result holds:
up to our numerical accuracy, all of the zeroes in a strip
along the real 
axis have imaginary part {\it exactly} equal to $\pm\pi/12$.
As $|r|=\rho\rightarrow 0$, the zeroes on the upper half line slide off
towards $+\infty$, while those on the lower head for
$-\infty$. The pattern becomes that of a pair of kink systems, 
one starting near $\theta=-\log(1/r)$, the other near
$\theta=+\log(1/r)$.
Conversely, as $\rho\rightarrow\rho_0$, the two sets
approach each other, although they always remain in their respective
left and right halves of the complex plane.

If $r$ is continued in a clockwise sense about the critical point and
back to the negative-$\lambda$ line, it
turns out that {\it none} of these zeroes cross the real axis. The
original TBA system (\ref{holst}) continues to hold, but its solution
$\ep(\theta)$ has nevertheless undergone a nontrivial monodromy.
The values of the scaling function which result form the upper 
set of points in figure~\ref{neglambdafig}, 
and match perfectly with the TCSA data for
the first excited state, at least in the range of $\rho$ for which
our rather crude iteration scheme is stable. Hence the 
TBA equation (\ref{holst}) is at least doubly-degenerate along the
negative-$\lambda$ line, with the second solution as physically
relevant as the first. The monodromy has an interesting effect on the
pattern of zeroes of $z(\theta)$: while their imaginary parts apparently
remain at $\pm\pi/12$, their real parts are no longer so simply
arranged, at least for $\rho$ smaller than about $2.8$. The rightmost
zero on the lower half line moves into the right half plane
$\re(\theta)>0$, thus lying to the right of the leftmost zero on the
upper half line -- the left and right kink systems have become
entwined, a feature that persists even as the two systems try to split
apart in the $\rho\rightarrow 0$ limit. On the other hand as $\rho$
increases past $2.8$, the ordering is briefly restored, though there
does not seem to be any great significance to this fact. Our numerics
for the upper branch quickly become unstable in this region, but an
extrapolation of the positions of the errant pair of zeroes is
consistent with their moving continuously to the same positions
(${}\approx\pm(0.24{+}\imath\pi/12)\,$) as found for the first two
zeroes on the lower branch, as $\rho\rightarrow\rho_0$. This supports
our supposition that the basic TBA system describes the first excited
state on the whole segment $0<\rho<\rho_0$, and not just on that part
where our iterations converged.

Having found the excited state for $r$ on the negative-$\lambda$ line, 
we now continue $r$ back to the real axis. As $\Arg(r)$ decreases from
$5\pi/12$, all of the zeroes of $z(\theta)$ bar the first on the upper
half line start to move up, towards the line $\im\theta=\pi/2$. The
first zero, on the other hand, is observed to move down, towards the
real axis. The zeroes on the lower half line behave in a symmetrical
fashion, as indeed they must given the $\theta\rightarrow -\theta$
symmetry of the basic equations. 
This is precisely the situation mentioned above, and we should be
ready to modify the TBA equations when the two singularities
actually hit the axis. Unfortunately the iterative solution of the
equation becomes unstable when singularities in $L(\theta)$ get too
close to the integration contour. It is possible to get a little
further by distorting the contour along which the equations are being
solved; in any event,  a Pad\'e extrapolation of the positions of the
two singularities under suspicion
clearly showed them crossing the real axis as $\Arg(r)$
decreased.
\setlength{\unitlength}{1.mm}
\begin{figure}[htbp]
\vspace{-1.5cm}
\hspace{1.cm}
\epsfxsize=9.5cm
\epsfysize=9.5cm
\epsfbox{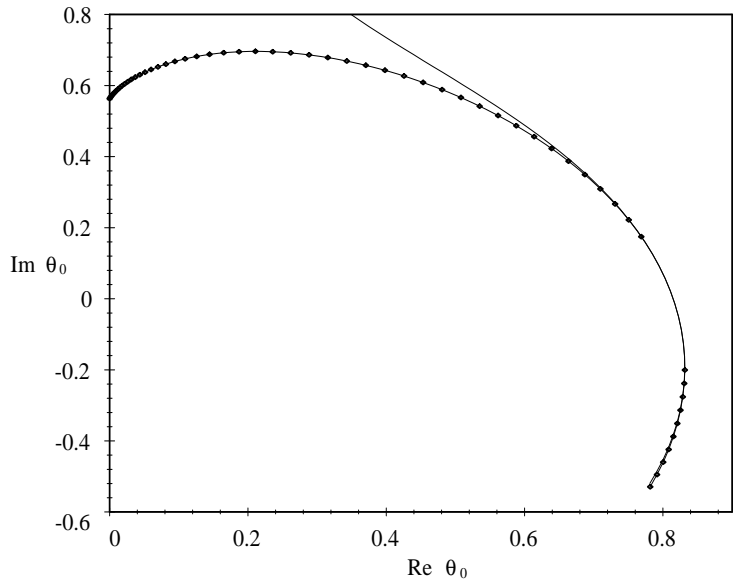}
\vspace{-0.9cm}
\caption{ \leftskip=1.3cm Pad\'e extrapolation of $\theta_0$ for the
first excited state, from $r=1.5\imath$ (lower right) to
$r=4.5$ (upper left), and back.
\rightskip=0.7cm }  
\label{padefig}
\end{figure}
Assume that $r$ is such that these
two singularities in $L(\theta)$ {\it have} crossed the real
axis, and now lie at $-\theta_0$, $\theta_0$. Here and in analogous
situations later on, we adopt
the convention that of the pair $\{-\theta_0,\theta_0\}$, it is
$\theta_0$ which has the positive imaginary part after the axis is
crossed.
Coming in from the left,
the integration contour for the convolution in (\ref{holst}) 
must now first loop down and
around the singularity at $-\theta_0$, and then back up and over the
singularity at $+\theta_0$, before proceeding to $+\infty$ along the
real axis.
An integration by parts turns these logarithmic singularities
into simple poles; evaluating the residues then
allows the equation to be recast with the contour running along the
real axis again:
\eq
\ep(\theta)=r\cosh\theta+\log{S(\theta-\theta_0)\over
S(\theta+\theta_0)}-\phi{*}L(\theta)~.
\label{cage}
\en
Similarly, the expression for $F_0(r)$ is modified, with $c_0$ of
equation (\ref{britten}) being replaced by
\eq
c(r)= {12r\over\pi}\imath\,\sinh\theta_0
 +{3\over \pi^2}\iintd\,
 r\cosh\theta L(\theta)~.~~
\label{parry}
\en
Although these equations appear to contain an unknown parameter,
namely $\theta_0$, this is not the case: self-consistency demands that
$\theta_0$ coincide with the position of the singularity in
$L(\theta)$ that necessitated its introduction in the first place. 
In this case, this translates as $\ep(\theta_0)=\imath\pi$ (the branch
of the logarithm to choose here can be fixed by 
continuity, starting from the
situation before the singularity crosses the axis). Substituting
$\theta=\theta_0$ into (\ref{cage}) then gives
\eq
0=r\cosh\theta_0-\log S(2\theta_0) - \phi{*}L(\theta_0)~.
\label{verdi}
\en
Flipping between equations (\ref{cage}) and (\ref{verdi}), 
iterative schemes can be set up which are convergent in
most regions of interest, given a reasonably accurate estimate of
$\theta_0$ for the initial iterate.
Starting with the extrapolated singularities of the
`excited' solution to (\ref{holst}), a solution to (\ref{parry})
can be picked up and followed all the way back to the real $r$ axis.
In this way, we tracked a solution along the line $r=t+(1.5{-}t/3)\imath$,
with $t$ varying from $0$ to $4.5$. Equation (\ref{holst}) converged
for $t$ less than about $0.45$, while (\ref{cage}) took
over as soon as $t$ became larger than $1$. The problems for
$0.45<t<1$ seem to be artifacts of our iteration schemes, and in
particular the good agreement the Pad\'e-extrapolated singularity
positions, both forwards and backwards, leave little doubt that the
gap was crossed correctly. Figure~\ref{padefig} matches these
extrapolations (shown by the continuous lines) with a selection of
points from the `raw' data. Points in the lower half plane (before
the singularity has crossed the axis) derive from the unmodified TBA
system~(\ref{holst}), and those in the upper half plane from the
modified system~(\ref{cage}). (For the backwards fit, where a greater
range of $t$ was available along which to collect data from which to
extrapolate, the matching is so good that the discrepancies with the
target points are rather hard to spot on the figure.)
It is also possible to monitor the behaviour of $c(r)$, finding good
agreement with TCSA data in both regimes.

As $r$ approaches the real axis, we see that $\theta_0$ approaches 
$\imath(\pi/6+\delta(r))$,
with $\delta(r)$ a small positive correction which tends to zero for large
real $r$. This information is enough to pin down the large-$r$
asymptotics of (\ref{verdi}). As $r$ grows, the convolution term becomes
relatively small, and the first two terms must cancel between
themselves. The only way this can happen as $r$ grows is for
$2\theta_0$ to approach a singularity of $S$; with
$\theta_0\sim\imath\pi/6$, the
singularity at $\imath\pi/3$ is the relevant one. Substituting 
$\theta_0(r)=\imath(\pi/6+\delta(r))$ and solving gives:
\eq
\theta_0(r)\sim\imath\lf(\pi/6+\sqrt{3}e^{-\sqrt{3}r/2}\ri)~.
\label{mahler}
\en
This immediately tells us the leading asymptotic of $c(r)$: substituting
(\ref{mahler}) into 
first term in (\ref{parry}) gives 
$c(r)\sim  -6 r (1 + 3 e^{-\sqrt{3}r/2}) / \pi$.
The next term is also easy to find, dropping the third factor in 
(\ref{cage}) and using the zeroth-order value of $\theta_0$,
namely $\imath\pi/6$, in order to find the leading behaviour of
$\ep(\theta)$ for substitution into (\ref{parry}). 
Gathering everything together we find
\eq
c(r)\sim { -6 r \over \pi} \lf(1 + 3 e^{-\sqrt{3}r/2}
 -{1\over 2\pi}\iintd\,
 \cosh\theta~ S(\theta+\imath \pi/ 2) e^{- r \cosh \theta } 
\ri)~.~~
\en
where the bootstrap relation $S(\theta+\imath \pi /6 )
/S(\theta-\imath \pi /6 ) =S(\theta+\imath \pi /2)$, relevant because
$\theta_0$ has been given its asymptotic value,
was used to reduce the
two S-matrices in (\ref{cage}) to one. This matches exactly with the
asymptotics predicted in refs~\cite{YZa,KMd} for the spin-zero
one-particle state, and therefore lends strong
support to our proposal. Further evidence will come from the numerical
comparisons with TCSA data to be reported shortly, but first we would
like to mention a natural generalisation of equations (\ref{cage}) --
(\ref{verdi}) which seems to capture all the remaining one-particle 
states.

The equations to consider read, for $r\in\RR$, as follows:
\bea
\ep(\theta)&=& r\cosh\theta +\log{S(\theta-\theta_0)\over
S(\theta-\bar\theta_0)} -\phi{*}L(\theta)~;\nn\\
c(r)&=&\imath {6r\over\pi}(\sinh\theta_0{-}\sinh\bar\theta_0)
+{3\over\pi^2}\iintd\,r\cosh\theta L(\theta)~,
\label{vivaldi}\\ \nn
\eea
where $\theta_0$, $\bar\theta_0$ are the complex-conjugate locations
of a pair of singularities in $L(\theta)$. We will discuss the
one-particle states with positive spin, and so we take $\re(\theta_0)>0$; 
the negative-spin states work similarly. The earlier equations at
real values of $r$
are recovered if $\theta_0$ is forced to be purely imaginary; conversely
if $r$ were to become complex in (\ref{vivaldi}), then $\theta_0$ and
$\bar\theta_0$ would generally
cease to be complex conjugates, and would have to
be tracked individually. Equation (\ref{verdi}) must also be modified,
both because of the changed form of (\ref{vivaldi}) and also to
allow for a more general singularity at $\theta_0$,
namely $\ep(\theta_0)=(2n{+}1)\pi\imath$ (and $n>0$ for
$\re(\theta_0)>0$). The equation becomes:
\eq
2n\pi\imath=r\cosh\theta_0-\log S(2\imath\im(\theta_0))-
\phi{*}L(\theta_0)~.
\label{maxwelld}
\en

Part of the motivation for these equations came from the form
of the two-particle equations to be introduced in the next section;
but to see immediately that they have a chance of being correct,
consider the large-$r$ asymptotics. The convolution term is sub-leading, 
and again it turns out that the balance between the first two terms is
achieved via $\im\theta_0=\pi/6+\delta(r)$ with $\delta(r)$ vanishing
as $r\rightarrow\infty$. Now take the imaginary part of
(\ref{maxwelld}) (previously this vanished automatically for real
$r$), and consider its behaviour as $r$ becomes large:
\bea
2n\pi&=& r\sinh\lf(\re(\theta_0)\ri)\sin(\pi/6{+}\delta(r))- \im\log
S(2\imath(\pi/6+\delta(r)))\nn\\
&\sim& {r\over 2}\sinh\lf(\re(\theta_0)\ri)
-\halft(1{-}{\rm sign}(\delta(r)))\pi~.
\label{mozart}\\ \nn
\eea
The term $\halft(1{-}{\rm sign}(\delta(r)))\pi$ is included to allow
for $\delta(r)$ being negative, in which case
$S(2\imath(\pi/6+\delta(r)))$ is negative and the logarithm picks up
an imaginary part.  Feeding this into (\ref{vivaldi}) gives
\eq
c(r)\sim  - {6 r \over \pi} \sqrt{1+(2\pi s/r)^2}
\label{smith}
\en
with $s= 2n + \halft(1{-}{\rm sign}(\delta(r)))$.
This is exactly as expected for a one-particle state with spin $s$.

One final observation: for all but the spin-zero one-particle state, 
equation (\ref{vivaldi}) is not symmetrical under $\theta\rightarrow
-\theta$. Hence it can never be obtained by analytic continuation of
the ground-state equation. But this is just as one would expect: the
entire Hilbert space of the SLYM splits up into sectors of different
spin, and analytic continuation can only ever move levels around within
a given sector. 

\setlength{\unitlength}{1.mm}
\begin{figure}[htbp]
\vspace{-1cm}
\hspace{1.cm}
\epsfxsize=9.5cm
\epsfysize=9.5cm
\epsfbox{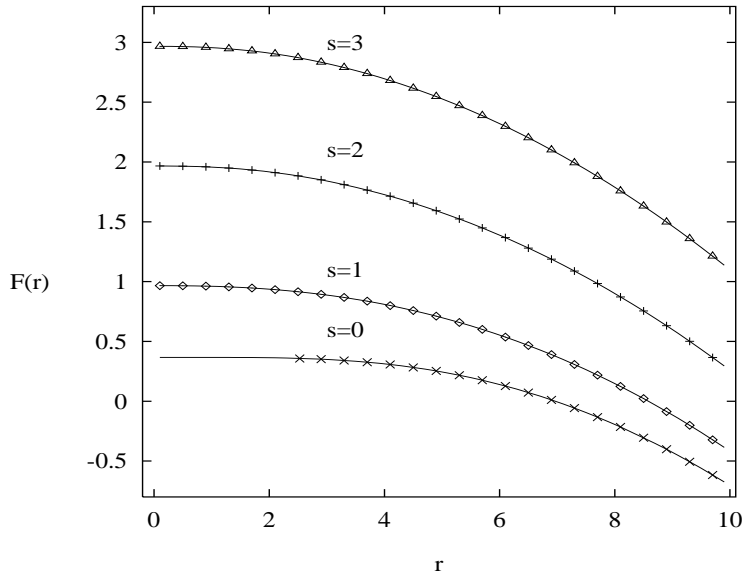}
\vspace{-0.8cm}
\caption{ \leftskip=1.3cm  Proposed one-particle scaling functions (points)
compared with TCSA data (continuous lines)   \rightskip=.7cm }  
\label{oneparticlefig}
\end{figure}

Figure~\ref{oneparticlefig} 
compares the numerical solutions of equations (\ref{cage}),
(\ref{maxwelld}) with TCSA data, for the first four one-particle
levels. We used the TCSA program of ref.~\cite{LMa}, 
truncating the quasiprimary
fields at level 5 and including as many derivative states as the
program allowed. The discrepancies 
are too small to see on the figure, being of order $10^{-9}$ at small
values of $r$, growing to $10^{-3}$ --- $10^{-4}$ (depending on the
level of the state) at $r{=}10$. The former
errors can be traced to our numerics and should not be too hard to
reduce, whilst the latter can be ascribed to truncation effects in the
TCSA -- in particular, they grow  as the level of the state under
consideration gets higher.

The only problem that the reader might notice is that the lowest set of
points stops short, at $r\approx 2.53$.
By this stage $\delta(r)$ has become so large that $\theta_0$ has almost
reached $\imath\pi/3$. We should expect
trouble at such a point as a new kind of singularity enters into the
equations, caused by singularities in $S(\theta{-}\theta_0)$
and $S(\theta-\bar\theta_0)$.  We have tried to take this into
account, but the resulting equation appears to be much less stable and
has resisted our attempts at an iterative solution. 
We will return to this problem in section~4; 
in any event it only appears to trouble the spin-zero
state.

\begin{figure}[htbp]
\vspace{-5.1cm}
\hspace{1.5cm}
\epsfxsize=13.cm
\epsfysize=9.5cm
\epsfbox{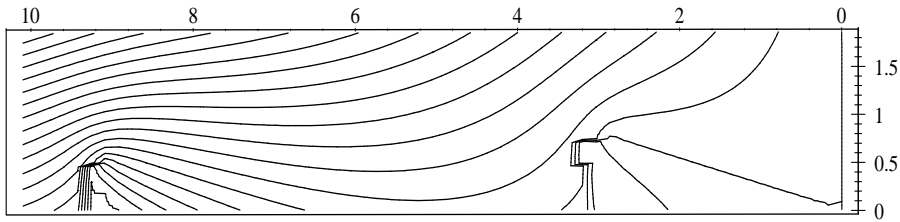}
\vspace{-0.5cm}
\caption{ \leftskip=0.75cm A contour plot of $\im(F(r))$ as obtained
from the basic TBA equations (1.5), (1.9), with $r$ lying in the box
$0 \le \re(r) \le 1.85$, $0 \le \im(r) \le 10.1$, showing the branch points
at $r_0$ and $r_1$, and also the segment of the negative-$\lambda$
line, running from $0$ to $r_0$, along which $\im(F(r))=0$.  The box
was scanned from right to left, so it is the lower branch of this 
segment that is visible.
\rightskip=0.55cm }  
\label{zeroplot}
\end{figure}

\resection{Two-particle states and beyond}
The singularity at $r_0=\rho_0e^{5\pi\imath/12}$ is only the first of a
whole sequence of singularities seen by the `zero-particle' TBA
(\ref{holst},\ref{britten}). They are approximately evenly-spaced along 
the direction of the imaginary axis: the next is at $r_1\approx
0.5311{+}9.1346\imath$, and the next at $r_2\approx 0.42{+}15.44\imath$. 
Figure~\ref{zeroplot} shows the positions of $r_0$ and $r_1$, as
emerged from the numerical solution of the basic TBA equations on a
suitably-fine grid.
\setlength{\unitlength}{1.mm}
\smallskip

On the analytical side, we note that
the first iterative correction to the Ising-like behaviour 
of (\ref{holst}), namely $\phi{*}\log(1{+}e^{-r\cosh\theta})$, tends
uniformly to zero as $\im(r)\rightarrow\pm\infty$ with $\re(r)>0$ held
fixed\footnote{when checking this, it is helpful to note that
$\int_0^{2\pi}\log(1{+}\alpha e^{\imath\gamma})d\gamma=0$ 
for $|\alpha|<1$ .}. Hence
we expect that the locations of the $r_n$ will eventually
approach 
$(2n{+}1)\pi\imath$ as $n\rightarrow\infty$. These appear to be the
only branch points on that part of the Riemann surface of $F(r)$ explored
by the zero-particle TBA. However the full surface must have much more
structure: the behaviour of the action~(\ref{arnold})
under $\lambda\rightarrow\bar\lambda$ means
that each singularity $r_n$ beyond the symmetrically-placed $r_0$ must
have an image on some other sheet, located at
\eq
\tilde r_n\equiv e^{5\pi\imath/6}\bar r_n~.
\label{debussy}
\en 
These images, invisible to the zero-particle TBA, should
instead be seen by the more general TBA systems that we are trying 
to construct.

\begin{figure}[htbp]
\vspace{-2cm}
\hspace{2.cm}
\epsfxsize=9.5cm
\epsfysize=9.5cm
\epsfbox{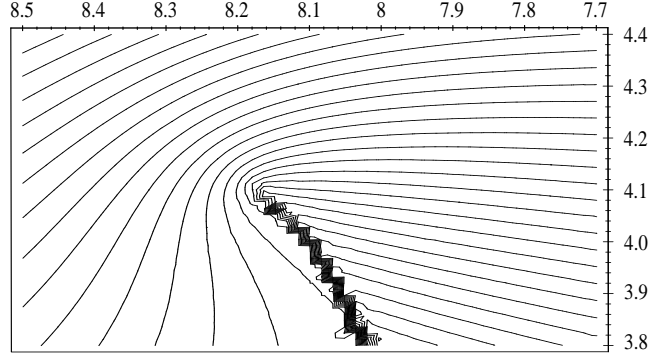}
\vspace{-1.2cm}
\caption{ \leftskip=0.6cm $\im(F(r))$ in the box $3.8 \le \re(r) \le 4.4$,
$7.7 \le \im(r) \le 8.5$ as obtained from the one-particle TBA equations
(2.3), (2.4), showing the branch point 
at $\tilde r_1$. \rightskip=0.55cm }  
\label{oneplot}
\end{figure}

This idea is confirmed by the values of $F(r)$
which result from the one-particle TBA introduced in the last
section: a grid plot of its real and imaginary parts
exhibits a square root singularity at 
$\tilde r_1\approx 4.1074{+}8.1763\imath$. An example of such a plot
is shown in figure~\ref{oneplot}.
(Incidentally, that this works
is a further piece of support for the results 
of the last section.) Judicious use of the TCSA, allied with
TBA data as a check on accuracy, is a great help in mapping out how
the various sheets fit together, and seems to be qualitatively
reliable at least out to $|r|{=}20$. 
Here, it tells us that the branch point at $\tilde r_1$ should
connect with the first two-particle state. We can therefore play the
same game as before, and study the behaviour of the solution
$\ep(\theta)$ of the one-particle TBA as $r$ is continued round
$\tilde r_1$. This should yield a two-particle TBA equation. 

\setlength{\unitlength}{1.mm}

When $r$ approaches the branch point, we found
that the singularities at $\theta=\pm\theta_0$, already implicated in
the one-particle equation, remained near $\pm\imath\pi/6$, whilst a second
pair headed for the real axis, just as happened when the
one-particle TBA was being formed. Strictly speaking we should now
repeat the rest of the one-particle work, this time keeping track of two
independent singularity positions, $\theta_0$ and $\theta_1$, as $r$
returns to the real axis. Numerically this is delicate:
with both $\theta_0$ and $\theta_1$ to locate, an efficient iteration
scheme is hard to find, and so we will leave this question to one side
for the time being.
Besides, in some respects the key piece of information has already
been obtained: the two-particle TBA equation should
involve four singularity terms, tied to singularities in $L(\theta)$
at $\pm\theta_0$ and $\pm\theta_1$. Once $r$ reaches the real axis,
the situation changes favourably: we expect the four singularities to
be invariant not only under $\theta_i\rightarrow -\theta_i$, but also
under $\theta_i\rightarrow\bar\theta_i$. This reduces to one the number 
of independent singularity positions, making a numerical solution no
harder than the cases examined in the last section. The TBA
equation to solve reads:
\bea
\ep(\theta)&=& r\cosh\theta +
 \log{S(\theta-\theta_0)\over S(\theta-\bar\theta_0)} 
+ \log{S(\theta+\bar \theta_0)\over S(\theta+\theta_0)} 
-\phi{*}L(\theta)~;\nn\\
c(r)&=&\imath {12 r\over\pi}(\sinh\theta_0{-}\sinh\bar\theta_0)
+{3\over\pi^2}\iintd\,r\cosh\theta L(\theta)~,
\label{ravel}\\ \nn
\eea
with $\theta_0$ satisfying
\eq
\ep(\theta_0) = (2n+1) \imath \pi
\label{ives}
\en
For the lowest two-particle state, found on continuing round $\tilde
r_1$, we found $\ep(\theta_0)=\imath \pi$, so that $n$ is equal to zero.
However, the results of the last section make it very natural to allow
for the more general possibility, in the hope of catching the other
two-particle states.  As with the one-particle TBA equation
(\ref{vivaldi}), $\theta_0$ and $\bar\theta_0$ cease to be complex
conjugate if $r$ strays from the real axis, eventually metamophosing
into $\theta_0$ and $-\theta_1$ as $\tilde r_1$ is
approached.

The analysis of the infrared  limit proceeds in much the same way
as for the $s\ne 0$ one-particle states. Substituting
$\theta=\theta_0$ in the first of~(\ref{ravel}) and taking the
imaginary part of the large-$r$ asymptotic gives us
\eq
2\pi(2 n +\halft(1{-}{\rm sign}(\delta(r))))  \sim 
r\sinh\lf(\re(\theta_0)\ri) + 
2\im\log{S(\theta_0+\bar \theta_0)\over S(2\theta_0)} 
\label{mozart2}
\en
where, as before, $\im\theta_0=\pi/6{+}\delta(r)$ and
$\delta(r)\rightarrow 0$ as $r\rightarrow\infty$. In the functions
$S$, $\delta(r)$ can be replaced by its limiting value, so
these terms become
\eq
\im \log S(2 \re(\theta_0)+ \imath {\pi/3}) = \halft \im \log 
S(2 \re(\theta_0)) \en
(using the $\phi^3$ property of $S(\theta)\,$) and
\eq
\im \log S(2 \re(\theta_0)) = - \imath \log S(2 \re(\theta_0))
\label{walton}
\en
(Recall that $S(\theta)$ is a pure phase for $\theta$ real, so the
right-hand side of~(\ref{walton}) is indeed real.)
Combining all of these terms together, (\ref{mozart2}) becomes
\eq
2\pi(2 n +\halft(1{-}{\rm sign}(\delta(r))))  \sim 
r\sinh\lf(\re(\theta_0)\ri) - \imath\log S(2\re(\theta_0))~.
\label{mozart3}
\en
This is just the Bethe-Ansatz quantization condition for a
two-particle state with rapidities $(-\re(\theta_0),\re(\theta_0))$ and
Bethe quantum numbers
\eq
\lf(-2n{+}\halft {\rm sign}(\delta) ,2n{-}\halft {\rm sign}(\delta) \ri)
=\lf(-{1 \over 2},{1 \over 2} \ri),\lf(-{3 \over 2},{3 \over 2} 
\ri),\lf(-{5 \over 2},{5 \over 2} \ri) \dots 
\label{purcell}
\en
with $n=0,1,1,2,2, \dots$ and ${\rm sign}(\delta) = -1,1,-1,\dots$. 
(cf, for example, equation~(4.5) of ref.~\cite{YZa}). These values of
${\rm sign}(\delta)$ should be imposed when handling the equation
numerically, so that when seeking to follow a particular solution, the
idea is to specify not only the value of $\ep(\theta_0)$,
equation~(\ref{ives}), but also the sign of $\delta(r)$. 
The only point where we were not able to impose a particular sign on
$\delta$ and obtain reasonable results was when we attempted to 
set ${\rm sign}(\delta)=+1$ when $n$ in
(\ref{ives}) is equal to zero. 
{}From a physical point of view, this is as it should be, given
the exclusion principle -- otherwise, the state $(-\halft,\halft)$
would have appeared twice.

\begin{figure}[htbp]
\vspace{-1.5cm}
\hspace{1.cm}
\epsfxsize=9.5cm
\epsfysize=9.5cm
\epsfbox{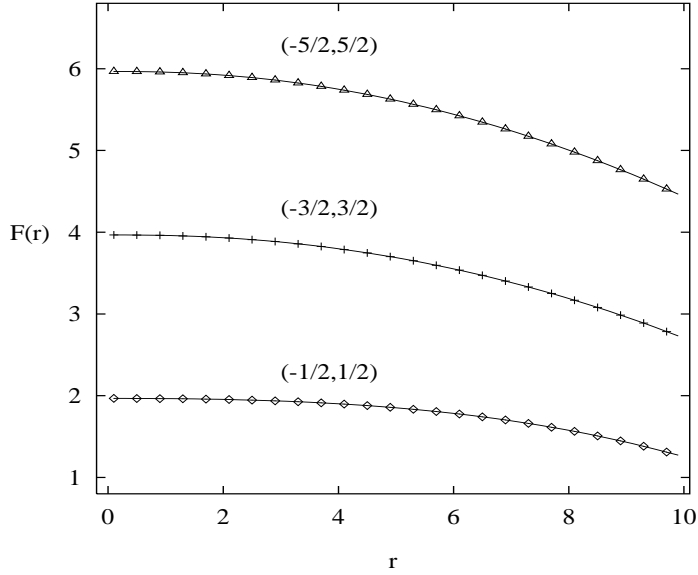}
\vspace{-0.3cm}
\caption{ \leftskip=1.3cm  Proposed zero-momentum two-particle scaling
functions 
(points) compared with TCSA data (continuous lines)  \rightskip=.7cm }  
\label{twoparticlesfig}
\end{figure}

Substituting the value of $\re(\theta_0)$ which follows from (\ref{purcell}),
together with $\im(\theta_0)=\pi/6$, into (\ref{ravel})
automatically gives the
expected algebraic asymptotic for $F(r)$. The next section will
examine the ultraviolet limits analytically; in the meantime 
figure~\ref{twoparticlesfig}
illustrates the numerical agreement with TCSA data for the first three
states. The discrepancies were much the same as for the one-particle
sector, described earlier.

\setlength{\unitlength}{1.mm}

By this point the structure is clear enough that we can conjecture a
general equation for an $n$-particle state:
\bea
\ep(\theta)&=& r\cosh\theta +
\sum_i \log{S(\theta-\theta_i)\over S(\theta-\bar\theta_i)} 
-\phi{*}L(\theta)~;\nn\\
c(r)&=&\imath \sum_i {6 r\over\pi}(\sinh\theta_i{-}\sinh\bar\theta_i)
+{3\over\pi^2}\iintd\,r\cosh\theta L(\theta)~,
\label{birt}\\ \nn
\eea
We expect that particular states are selected by imposing the values
of $\ep(\theta_i)$, and also the signs of the various $\delta_i$. We
only analysed in detail the spin zero three-particle states, where 
steps similar to those already described give the following
Bethe Ansatz quantum numbers:
\eq
(-2n+\halft+\halft{\rm sign}(\delta),0,2n-\halft-\halft{\rm sign}(\delta))
=(-1,0,1),(-2,0,2),\dots
\en
with $n=1,1,2,2,3...$ and ${\rm sign}(\delta)=+1,-1,+1,\dots\,$.
Again, the exclusion principle leads us to suppose that the $n{=}0$
case should be omitted. 
 
This large-$r$ asymptotic, together with the special cases already studied 
in some detail, lend support to (\ref{birt}). However numerical work
will also be needed. In particular,
the equation will have to be modified at values of $r$ below some
critical $r_c$ whenever the phenomenon observed above for the
spin-zero one-particle state occurs, and the imaginary part of 
a singularity position $\theta_i$ ventures too far from $\pi/6$. 
We suspect that this will happen whenever one of the Bethe
Ansatz quantum numbers is equal to zero, and thus will
afflict the three-particle states just discussed. In fact there are
signs that this is just what is needed to get the ultraviolet
asymptotics correct in these cases, a point that we shall return to
later.

\resection{Ultraviolet behaviour}
Analytic work has thus far been restricted to the infrared regime.
In this section the opposite limit is considered, and the ultraviolet
scaling dimensions of various states extracted. We start with the
two-particle states.

In the limit 
$r \rightarrow 0$, the two-particle TBA splits into a pair of kink
systems, just as for the ground state. For the right kink system, the
terms from the singularities at $-\theta_0$ and $-\bar\theta_0$ drop
out, and $r\cosh\theta$ can be replaced by $\halft re^{\theta}$. 
The real part of $\theta_0$ tends to infinity like
$\log(1/r)$; we will also need a little information about the
behaviour of the imaginary part. Recall that for large $r$,
$\im(\theta_0)\sim\pi/6{+}\delta(r)$ with $\delta(r)\rightarrow 0$ as
$r\rightarrow\infty$. In the opposite direction, we find that,
although $\delta(r)$ grows, it tends to a finite and still-small
limit as $r\rightarrow 0$. The precise value drops out of the final
equations; the only important fact is that for all of the two-particle
states (and also for all of the $s{\ne}0$ one-particle states) its
absolute value is less than $\pi/6$. Quite why this is important
should become clear shortly. The scale-invariant kink TBA reads:
\eq
\ep(\theta)= {r\over 2}  e^{\theta}
+\log { S(\theta-\theta_0) \over  S(\theta-\bar{\theta}_0)}
- \phi{*}L(\theta)~.
\label{grainger}
\en
(Note, to pass from this `chiral' TBA equation to the general
one-particle equation introduced in section~2, 
we just need to replace $\halft
e^{\theta}$ with $\cosh\theta$.) Next, eliminate $r$ by replacing
$\theta$ with $\theta-\log r$ and $\ep(\theta)$ by $\ep(\theta-\log
r)$, to find
\eq
\ep(\theta)= \halft e^{\theta}
+\log { S(\theta-\eta) \over  S(\theta-\bar{\eta})}
- \phi{*}L(\theta)~,
\label{puccini}
\en
where $\eta=\theta_0{+}\log r$. In terms of these quantities, the
limiting value of $c(r)$ is
\eq
c= {3 \over \pi^2} \iintd\, e^{\theta} L(\theta)
+{6\over\pi}\,\imath\,(e^{\eta} -e^{\bar\eta} )~.
\en
Now we proceed in the usual manner. First take a derivative with
respect to $\theta$:
\eq
\prtial\ep(\theta)=\halft e^{\theta}
+\prtial \log {S(\theta-\eta) \over 
S(\theta-\bar\eta)} 
-\prtial \phi{*}L(\theta) ~.
\en
Now substitute this into the formula for $c$:
\eq
c= {6 \over \pi^2}  \iintd L(\theta) 
\prtial\lf(\ep(\theta) - 
\log {S(\theta-\eta) \over S(\theta- \bar\eta)}
+\phi{*}L(\theta)\ri) 
 {}+{ 6\over\pi}\,\imath\,(e^{\eta}-e^{\bar\eta})~.~~ 
\en
Then (remembering that $\phi(\theta)=-\imath\prtial\log S(\theta)$)
\bea
c&=& {6 \over \pi^2}\lf[ \int_{\ep_{min}}^{\ep_{max}}dx\,\log(1+e^{-x}) + 
{1 \over 2} \ep_{min}\log(1+e^{-\ep_{min}}) \ri]  \nn \\
&&{}-{6\over\pi}\,\imath\,\lf[ 
2\phi{*}L(\eta)-2\phi{*}L(\bar\eta)-e^{\eta}+e^{\bar\eta} \ri]~.
\eea
The first piece can be expressed in terms of the Rogers' dilogarithm 
function ${\cal L}(z)=-\halft\int_0^zdt\lf({\log(1-t)\over t}+{\log
t\over 1-t}\ri)$ as
\eq
{6\over\pi^2}\lf[{\cal L}\lf({1\over 1+e^{\ep_{min}}}\ri)
 -{\cal L}\lf({1\over 1+e^{\ep_{max}}}\ri)\ri]~,
\en
whilst the second
can be evaluated using (\ref{puccini}) for $\eta$ and
$\bar\eta$:
\bea
2n \pi \imath &=& 
\halft e^{\eta}-\log(|S(\imath 2 \im(\eta))|)
              - \pi \imath \halft(1{-}{\rm sign}(\delta) )
              -\phi{*}L(\eta) \nn\\
-2n \pi \imath &=& \halft e^{\bar\eta}+\log(|S(\imath 2 \im(\bar\eta))|)
              + \pi \imath \halft(1{-}{\rm sign}(\delta) )
              -\phi{*}L(\bar\eta)~,~~ 
\eea
and is equal to $12(4n{+}1{-}{\rm sign}(\delta))\,$. 

Finally,  $\ep_{min}$ is reached as $\theta \rightarrow 
-\infty$, $\ep_{max}$ as $\theta\rightarrow\infty$. Since 
$S(\pm\infty)=1$,  the limiting form of (\ref{grainger})
gives $\ep_{min}=\log(1{+}\sqrt{5})/2)$ and $\ep_{max}=\infty$, 
a result which was also
checked against our numerical solutions.
Since
${\cal L}(2/(3{+}\sqrt{5}))=\pi^2/15$, this gives us
\eq
c = {2 \over 5} - 12 (4 n + 1 - { \rm sign}(\delta) )~.
\en
The calculation for the $s\ne 0$ one-particle states is essentially
identical. The only difference is that for one of the two kink systems
is in its `ground state', in that 
there are none of the extra terms involving $\theta_0$. Hence the size
of the additional contribution is halved, and
\eq
c = {2 \over 5} - 6 (4 n + 1- {\rm sign}(\delta) )~.
\en
Recalling from the infrared result~(\ref{smith}) that
$s=2n{+}\halft(1{-}{\rm sign}(\delta))$, this is $-1/12$ times the
scaling dimension on the cylinder of a spin $s$ descendant of the primary
field $\varphi$, as expected.

For $s=0$ in the one-particle sector, the situation changes and
our discussion must be much more
tentative, as our numerical work lost stability below $r=r_c\approx 2.53$.
Nevertheless we can conjecture a plausible modification to the
equation which seems to predict the correct ultraviolet asymptotic,
and the remainder of this section is devoted to this question.

First, we must decide just how we expect the equation to change. 
In particular  is important to know if the singularities 
at $\pm\theta_0$ stay on the imaginary axis, 
or if they somehow each split in two and acquire a real part. 
Unlikely as it appears at first sight, it seems that it is the latter
possibility which actually occurs.  One can gain information about
this by again returning $r$ into the complex plane, 
and  following the position of $\theta_0$ as $r$ returns to points
on the real axis below the critical value $r_c$. 
Pad\'e extrapolation clearly showed a finite limiting value of 
$\re(\theta_0)$ in each case we tried, 
the value increasing as the limiting $r$ decreased.
For example, extrapolating down from the upper half plane,
we obtained $\re(\theta_0)\approx 0.77$ at $r{=}1.5$, and 
$\re(\theta_0)\approx 1.17$ at $r{=}1$.
The extrapolation of $\im(\theta_0)$ was not so reliable, but the
results were consistent with the limiting value being $\pi/3$ every
time. In any event, this seems to be the only way to make sense of the
non-zero limit for the real parts, and so we will proceed on the
assumption that indeed $\im(\theta_0(r))=\pi/3$ for all $r<r_c$.

When $\im(\theta_0)$ finally reaches $\pi/3$, the singularities in
$L(\theta)$ caused by the $\log S(\theta\pm\theta_0)$ term
hit the real axis, forcing a deformation of the 
contour of integration.
Integrating by parts in the convolution, and then taking the principal
part to restore the contour to its original track,
equations~(\ref{cage}), (\ref{parry}) should be modified thus:
\bea
\ep(\theta)&=& r\cosh\theta 
+\log{S(\theta-\theta_0)\over
      S(\theta+\theta_0)} 
+ \halft \log{S(\theta+(\theta_0-\imath \pi/3))\over
      S(\theta-(\theta_0-\imath \pi/3))} 
-\phi{*}L(\theta)~;\nn\\
c(r)&=&
\imath {12 r\over\pi}\sinh\theta_0
-\imath {6 r\over\pi}\sinh(\theta_0 - \imath \pi/3)~\nn \\
&&{}~~~{}+{3\over\pi^2}\iintd\,r\cosh\theta L(\theta)~,
\label{vivaldi2}\\ \nn
\eea
The factors of $1/2$ appear because the singularities remain on the real 
axis. This equation looks rather complicated, but simplifies once we
put  $\theta_0=\beta{+}\imath\pi/3$, with $\beta$ real.  Using 
the bootstrap equation we find
\bea
\ep(\theta)&=& r\cosh\theta 
+ \halft \log{S(\theta{-}\beta{-} \imath \pi/3)\over
              S(\theta{-}\beta{+}\imath \pi/3)} 
+ \halft \log{S(\theta{+}\beta{-}\imath \pi/3)\over
              S(\theta{+}\beta{+}\imath \pi/3)} 
-\phi{*}L(\theta)~;\nn\\
c(r) &=&-{6 r \over \pi}\sqrt{3}\cosh\beta +
     {3\over\pi^2}\iintd\,r\cosh\theta L(\theta)~.
\label{vivaldi1}\\ \nn
\eea
To understand what has happened, it is worth thinking 
about how the `singular values' of $\ep(\theta)$ have moved around.
This can be discussed using the functions $Y(\theta)\equiv
e^{\ep(\theta)}$ which solve, for the Lee-Yang model, the following
functional equation or $Y$-system~\cite{Zf}
\eq
Y(\theta{-}\imath{\pi\over 3})\,Y(\theta{+}\imath{\pi\over 3})
=1+Y(\theta)~.
\en
The special points that we are interested in, $\theta_0$ being an
example, are those for which $1{+}Y=0$.  Given such a point
$\theta^{(0)}$, the $Y$-system can be used
to find a sequence of other points 
$\theta^{(k)}\equiv\theta^{(0)}{+}k\pi\imath/3$ 
where $Y$ also takes special values. Setting $Y^{(k)}=Y(\theta^{(k)})$,
with $Y^{(k{+}5)}=Y^{(k)}$ by the periodicity of the $Y$-system,
the sequence cycles round as $-1,A,B,-1,0$, with $A{+}B=0$ and
$Y^{(0)}$ either the first or the fourth term. For
$\theta_0$, the former option is realised and so as $\theta_0$ moves
around, it carries with it the points $\theta_0{-}\imath\pi/3$ and
$\theta_0{-}2\imath\pi/3$, at which $Y$ takes the values $0$ and $-1$
respectively. While $r>r_c$, these points lie on the imaginary axis,
together with a symmetrically-placed triplet of points associated with
$-\theta_0$.  When $r$ reaches $r_c$,
the points $\theta_0{-}2\imath\pi/3$ and $-\theta_0$ collide at
$-\imath\pi/3$. Since for both of these points
$1{+}Y=0$, this opens up another possibility to
distort the singularity positions while maintaining the symmetry
under $\theta\rightarrow\bar\theta$ that they must respect
while $r$ remains
real. This seems to be realised here: as $r$ decreases below
$r_c$, the points $\theta_0-2\imath\pi/3$ and $-\theta_0$
move away parallel to the real
axis, with a similar story at $+\imath\pi/3$ ensuring that the
symmetry is preserved. By continuing $r$ away through complex values,
we were able to observe this end result without having to tangle with the
particularly singular behaviour actually at $r{=}r_c$.

The last task is to find an equation for $\theta_0$, or 
equivalently for $\beta$. 
The value of $\ep(\theta_0)$ remains equal to $\imath\pi$, 
but one must be careful when substituting $\theta{=}\theta_0$ into the
first of
(\ref{vivaldi1}). This is because the singularities just discussed 
cause the right hand side of this equation to develop a couple of
singularities precisely when $\theta{=}\theta_0$. 
Of course, the overall result remains
finite, and the simplest approach seems to be to consider the limit of
$\ep(\beta{+}\imath\pi/3{-}\imath\epsilon)-
\ep(\beta{-}\imath\pi/3{+}\imath\epsilon)$
as $\epsilon\rightarrow 0^+$. From one point of view this is
equal to $2\imath\pi$; equating this with the result from the right
hand sides we found
\bea
\imath\pi & = & \imath \sqrt{3}\,r\,\sinh\beta
+  {1 \over 2} \log{S(2\beta)\over
                  S(2 \beta+\imath 2\pi/3)}  
                 {S(2\beta)\over
                  S(2\beta-\imath 2\pi/3)} \nn \\
&&{}- ( \phi{*}L(\beta+\imath \pi/3)-\phi{*}L(\beta-\imath \pi/3))~.
\eea
(We should mention here that we have been somewhat cavalier throughout
in our treatment of the branch choices for the logarithms. This issue
deserves a careful study, especially in regard to the way the branches
behave under analytic continuation between energy levels.)

This concludes the modifications to the one-particle equation. If the
comment made at the end of section~3 is correct, then a similar
man\oe uvre will be needed whenever a zero-momentum particle is present
in a Bethe Ansatz state.
The natural generalisation of the equations just obtained for such
situations, correcting equation (\ref{birt}) for $r$ less than some
$r_c$, is 
\bea \ep(\theta)&=& r\cosh\theta 
+ \halft \log{S(\theta-\beta- \imath \pi/3)\over
              S(\theta-\beta+\imath \pi/3)} 
              {S(\theta+\beta-\imath \pi/3)\over
              S(\theta+\beta+\imath \pi/3)} \nn \\
&&{}+\sum_i \log{S(\theta-\theta_i)\over S(\theta-\bar\theta_i)}
            {S(\theta+\bar\theta_i)\over S(\theta+\theta_i)}
 -\phi{*}L(\theta)~;\nn\\
c(r)&=& -{6 r\over\pi}\sqrt{3}\cosh\beta\nn\\
&&{}+\imath\sum_i{12r\over\pi}(\sinh\theta_i{-}\sinh\bar\theta_i)
+{3\over\pi^2}\iintd\,r\cosh\theta L(\theta)~,~~
\label{vivaldi5}\\ \nn
\eea
where as above the singularity position $\theta_0$ ceased to be purely
imaginary at $r{=}r_c$, 
and was replaced in the equations by the real variable
$\beta=\theta_0{-}\imath\pi/3$.

Now we must analyse the $r\rightarrow 0$ limit. The
first thing that we notice is that the modification has achieved the
remarkable trick of splitting the `zero-momentum' singularity position
$\theta_0$ into constituent singularities at $\imath\pi/3\pm\beta$,
which can now join the respective left and right kink systems.
For the one-particle case, assuming that $\beta\rightarrow\infty$
as $r\rightarrow 0$, and
replacing $\theta$ with $\theta-\log r$ and $\ep(\theta)$ with
$\ep(\theta-\log r)\,$, the right hand system becomes
\bea
\ep(\theta)&=& \halft  e^{\theta} 
+ \halft \log{S(\theta-\eta- \imath \pi/3)\over
              S(\theta-\eta+\imath \pi/3)} 
-\phi{*}L(\theta)~;\nn\\
c&=&
-{3 \over\pi}\sqrt{3}\,e^{\eta} +
{3\over\pi^2}\iintd\, e^{\theta} L(\theta)~,
\label{vival}\\ \nn
\eea
where $\eta=\beta+\log r$ satisfies the `quantization condition'
\eq
\imath\pi = \imath\,{\sqrt{3}\over 2}\,e^{\eta}
- \lf(\phi{*}L(\eta+\imath \pi/3)-\phi{*}L(\eta-\imath \pi/3)\ri)~.
\label{faure}
\en
Now the calculation runs as before, modulo one subtlety to be mentioned
shortly. One finds
\bea
c&=& -{3\sqrt{3}\over\pi}\,\,e^{\eta}+
  {6 \over \pi^2}\lf[ \int_{\ep_{min}}^{\ep_{max}}dx\,\log(1+e^{-x}) +
\halft\ep_{min}\log(1+e^{-\ep_{min}}) \ri]  \nn \\
&&{}-   
{6\over \pi}\,\imath\,\lf[\phi{*}L(\eta{+}\imath\pi/3)-
\phi{*}L(\eta{-}\imath\pi/3)\ri]~. 
\eea
Recognising the dilogarithm and  using (\ref{faure}), 
\eq
c= {6 \over \pi^2}\lf[{\cal L}\lf({1\over 1{+}e^{\ep_{min}}}\ri) - 
{\cal L}\lf({1\over 1{+}e^{\ep_{max}}}\ri) \ri] - 6~.
\en
As before, $\ep_{max}=\infty$, and $e^{\ep_{min}}$ solves 
$e^{2 \ep_{min}}=1+e^{\ep_{min}}$, an equation which has two solutions:
\eq
e^{\ep_{\pm}}={1 \pm \sqrt{5} \over 2}
\en
Previously we selected the positive solution, $\ep(\theta)$ being a
monotonically increasing real function in that case. 
However this time note that 
$e^{\ep(\infty)}=e^{\ep_{max}}=\infty$, and that as $\theta$ decreases
along the real axis
$e^{\ep(\theta)}$ falls all the way down to zero at
$\theta=\beta$. This suggests that the relevant solution by the time
$\theta=-\infty$ is reached is the negative one (note, such solutions
were previously observed to be of
relevance to the excited
states in some other models by Martins, in ref.~\cite{Mc}).
The correct prescription for  ${\cal L}(x)$ for general $x\in\RR^+$ 
is (see for example~\cite{Ka})
\eq
{\cal L}(x)={\pi^2\over 3}-{\cal L}(1/x)~~~{\rm for}~~x>1~,
\en
which in this case gives
\eq
{\cal L}(2/(3{-}\sqrt{5}))={\pi^2\over 3}
-{\cal L}(2/(3{+}\sqrt{5}))= {4\pi^2\over 15}~.
\en
This gives us $c=-22/5$, as expected for the spin-zero one-particle
state, which in the ultraviolet is created by the identity operator
$I$. (Note, the 
theory being non-unitary, this is {\it not} the ground state!)
We also analysed the spin zero 3-particle  sector, and found
\eq
c=-{22 \over 5} -12(4n{+}1{-}{\rm sign}(\delta))
\en
with $n=1,1,2,2,\dots$ and ${\rm sign}(\delta)=+1,-1,+1,\dots\,$ just as
in the infrared. Note that here also $n$ must
start from $1$: the two options at $n{=}0$ give either $-22/5$,
double-counting the state $I$, or $-22/5-24$, the would-be scaling
dimension on the cylinder of the null field $\partial\bar\partial I$.
This seems to suggest a relation
between the exclusion principle on the Bethe Ansatz quantum numbers, 
and the null field structure of the conformal states.

%

\resection{Conclusions}
This work is still in its early stages. We would like to have a more
secure understanding of the zero-momentum one-particle state, and in
particular to be able to follow its behaviour numerically all the way
down to $r{=}0$. Work on this question is in progress. The situation
for the remaining one-particle states, and for all of the two-particle
states, is very satisfactory but beyond that our equations become more
conjectural, albeit natural. Numerical and analytic work is needed,
both to confirm their status and to unravel their structure. On the
numerical side this poses the particular challenge of developing
efficient methods for the tracking of a number of singularities
simultaneously, but this should not be insurmountable -- the
three-particle state would be a good starting point. Quite apart from
the analytic insights we can hope for, the method promises to be very
competitive numerically with the TCSA, particularly for higher levels
and larger values of $r$. Note also that all of the results we have
obtained for the SLYM are directly relevant, after a multiplication
by two, to the thermally-perturbed three-state Potts model. This follows
from the simple relationship between their respective ground-state TBA
systems~\cite{Zb}.

Turning to more general issues, many new features of the TBA equations
seem to emerge when the whole complex $r$ plane is considered, and
there remains much to explore. The map
$\lambda\rightarrow\bar\lambda$, simple in its effects from the point
of view of the perturbative action~(\ref{arnold}), is far from trivial
when acting through equation~(\ref{debussy}), as 
$r\rightarrow\tilde r$, on the space of multiparticle
TBA equations. More locally in this space, the study of the zero, one
and two particle TBA systems has shown how one equation can
melt into another with the passage of singularities across the real
axis, and it is important to extend this treatment further. An
immediate hope would be to thereby justify the more general 
equations~(\ref{birt}) conjectured above. More ambitiously, given the
good understanding of the ultraviolet and infrared limits of
the model provided by conformal field theory and the Bethe-Ansatz 
quantization conditions respectively, one might try to say 
something about the
structure of the full Riemann surfaces for $F(r)$ in the various
sectors, and to understand the way in which the domains of the various 
TBA equations are patched together on these surfaces.

We would like to stress how general the method advocated in
this paper should be. As a first step, there seems to be no serious
obstacle to its application to the known purely elastic scattering
theories.  In the absence of any numerical work to report at this
stage, we will restrict ourselves to a couple of comments. First, we
note the crucial r\^ole that the $\phi^3$ property of $S(\theta)$
played in much of the analysis presented above. 
One can anticipate that more general cases will exhibit a similar
interplay between the algebraic properties of the $S$-matrix and the
asymptotics of the multiparticle TBA equations.
Second, many scattering theories, purely elastic and otherwise, 
exhibit additional symmetries which on the one hand
divide the Hilbert space into further subsectors,
and on the other permit the construction of 
alternative `seed' TBA~\cite{Fc}. These will provide 
additional starting-points for the continuation process, thus helping
to fill out the extra sectors. In the
Ising model example discussed in section~1, this was exactly how the
states related to the spin field arose.

Our main conclusion, however, is rather more general than this. 
It is that {\it any} ground-state TBA equation encodes within itself
equations for many excited states, and that analytic continuation
provides the means by which these equations can be extracted.
It will be interesting to see just how far this programme can be
carried through.

%
%
\vskip 0.4cm
\noindent
{\bf Acknowledgements} -- We would like to thank Francesco Ravanini
for some very useful discussions in the early stages of this project.
We are also grateful for comments and help at various points from 
Carl Bender, Elena Boglione, Kevin Thompson and Mathew Penrose.
PED thanks the EPSRC for 
an Advanced Fellowship, and
RT thanks the Mathematics Department of Durham University for a
postdoctoral fellowship.
This work was supported in part by a Human
Capital and Mobility grant of the European Union, contract number
ERBCHRXCT920069, and in part by a NATO grant, number CRG950751.

\vskip 0.4cm
\noindent
{\bf Note} -- Ref.~\cite{BLZa}, by Bazhanov {\it et al}, appeared as
we were finishing this paper. These authors independently
obtained TBA-like equations for the spin-zero excited states of the
SLYM, though by a very different route from the one that we have been
describing here.
%
%

\end{document}